\documentclass[aps,prb,twocolumn,showpacs,amsmath,amssymb,superscriptaddress]{revtex4-1}
\usepackage{graphicx}
\usepackage{dcolumn}
\usepackage{epstopdf}
\usepackage{bm}
\usepackage{color}

\begin{document}
\title{Interfacial interactions between local defects in amorphous SiO$_2$ and supported graphene}

\author{A.~N. Rudenko}
\email[]{rudenko@tu-harburg.de}
\author{F.~J. Keil}
\affiliation{Institute of Chemical Reaction Engineering, Hamburg University of Technology, Eissendorfer Strasse 38, D-21073 Hamburg, Germany}
\author{M.~I. Katsnelson}
\affiliation{Institute for Molecules and Materials, Radboud University Nijmegen, Heijendaalseweg 135, 6525 AJ Nijmegen, The Netherlands}
\author{A.~I. Lichtenstein}
\affiliation{Institute of Theoretical Physics, University of Hamburg,
Jungiusstrasse 9, D-20355 Hamburg, Germany}
\date{\today}

\begin{abstract}
We present a density functional study of graphene adhesion on a realistic SiO$_2$ surface taking into account van der Waals (vdW) interactions.
The SiO$_2$ substrate is modeled at the local scale by using two main types of surface defects, typical for amorphous silica: the oxygen dangling bond and three-coordinated silicon. 
The results show that the nature of adhesion between graphene and its substrate is qualitatively dependent on the surface defect type. In particular, the interaction between graphene
and silicon-terminated SiO$_2$ originates exclusively from the vdW interaction, whereas the oxygen-terminated surface provides additional
ionic contribution to the binding arising from interfacial charge transfer ($p$-type doping of graphene). Strong doping contrast for the different surface
terminations provides a mechanism for the charge inhomogeneity of graphene on amorphous SiO$_2$ observed in experiments.
We found that independent of the considered surface morphologies, the typical electronic structure of graphene in the vicinity of the Dirac point 
remains unaltered in contact with the SiO$_2$ substrate, which points to the absence of the covalent interactions between graphene and amorphous silica.
The case of hydrogen-passivated SiO$_2$ surfaces is also examined. In this situation, the binding with graphene is practically independent of the type of 
surface defects and arises, as expected, from the vdW interactions.
Finally, the interface distances obtained are shown to be in good agreement with recent experimental studies.
\end{abstract}

\maketitle

\section{Introduction}

Graphene, a single layer of graphite, is subject of intensive research
owing to its remarkable electronic properties, which make this material a promising 
candidate for various electronic applications.\cite{Novoselov,Geim,Graphene-RMP}
Since graphene is one-atom thick, its properties are strongly influenced
by the environment, such as, for example, atomic impurities,\cite{imp_Wehling} molecular 
adsorbates,\cite{halogen} metallic contacts,\cite{Khomyakov,Hamada} and dielectric 
substrates.\cite{sio2_Wehling,mica} In respect to practical applications, the
behavior of graphene in contact with substrates is of great importance.
Since the discovery of graphene the role of substrates in its properties is 
actively debated, but is still not clearly understood. 

Silicon dioxide is the main insulating component of present microelectronic devices 
owing to its perfect dielectric behavior and chemical stability.\cite{sio2_app} 
However, this material is not the best candidate for its use as a substrate for graphene-based devices. 
In comparison to the remarkably high carrier mobility observed in suspended graphene,\cite{Bolotin}
carrier mobility in graphene supported on SiO$_2$ is very limited by scattering from surface
states, by surface phonons, and by structural deformations of graphene.\cite{Chen,Chen2,Lui} The deformations are attributed not to 
intrinsic corrugations of graphene,\cite{Meyer,Fasolino} which might also exist in SiO$_2$-supported graphene,\cite{Geringer}
but rather to the morphology of the SiO$_2$ surface 
itself, which is amorphous and, therefore highly disordered.\cite{Ishigami} It has been demonstrated that
the rippling of graphene can be considerably suppressed by its deposition on relatively flat surfaces, such as,
for example, muscovite mica\cite{Lui} or hexagonal boron nitride.\cite{Dean}
Furthermore, it has been shown that the adhesion of graphene on SiO$_2$ is essentially conformal, meaning that graphene 
reproduces the substrate topography with very high accuracy.\cite{Cullen} Apparently, interfacial interactions play a key role in
the understanding of the structural behavior of graphene supported on surfaces. Nevertheless, the microscopic nature of the graphene 
adhesion on realistic SiO$_2$ has not yet been clearly established.

Apart from the structural changes of graphene, a substrate can directly affect its electronic properties by induced charges as shown by scanning tunneling microscopy (STM) experiments.\cite{Zhang} 
Furthermore, recent electrical field microscopy (EFM) experiments 
have demonstrated that the electrical potential of amorphous SiO$_2$ surface is rather disordered, which suggests the existence of localized chargelike 
impurities or surface defects.\cite{Garcia} The morphology of possible surface defects, their influence on the properties of graphene, and the bonding mechanisms in the 
graphene-substrate system are yet unclear from the theoretical point of view and require a detailed microscopic analysis.

So far, a number of papers reported theoretical first-principles investigations of graphene on SiO$_2$.\cite{sio2_Wehling,Kang,Shemella,Li,Hossain,Jadaun,Cuong} For modeling of the amorphous silica 
surface, various crystalline approximations are usually used paying not much attention to the structural peculiarities of realistic SiO$_2$. As a consequence, the results substantially differ from 
one approximation to another. For instance, the oxygen-terminated (001) surface of $\alpha$-quartz was predicted to be very reactive,\cite{Kang,Shemella} significantly affecting the electronic 
properties of graphene, whereas the corresponding reconstructed surface is essentially inert and has no influence on the electronic structure of graphene.\cite{Cuong} Although the structure of amorphous
SiO$_2$ is not trivial, it has only certain types of structural features (defects),\cite{Feuston,walsh_1} and thus could be rationally modeled only within the frame of specific crystalline approximations.
In addition, previous investigations of graphene on SiO$_2$ do not explicitly consider dispersive interactions, but employ only local density approximations (LDA) for the treatment of exchange-correlation 
effects.
Although the LDA interlayer distances are very close to experimental values for some layered 
materials (e.g., graphite), this approximation is purely local and not able to describe dispersive
interactions of nonlocal nature. As a consequence, it is impossible to reproduce accurately the binding
energies and elastic properties of vdW complexes within the LDA. As has been shown, the dispersive 
interactions play a significant role in the adhesion of graphene on silicate surfaces.\cite{mica} 
For this reason, it is desirable to use more sophisticated approaches, allowing for explicit treatment 
of the vdW interactions.

In this paper we present a first-principles van der Waals density-functional study (vdW-DF) of graphene supported on amorphous SiO$_2$. Particularly, we are focused on
the interaction between graphene and its substrate as well as on the electronic structure of supported graphene. 
We model the SiO$_2$ substrate by considering idealized surfaces with two main surface defects, typical for amorphous silica: the oxygen dangling bond and three-coordinated silicon.
It is assumed that this structural model represents a realistic SiO$_2$ surface at the local scale.
Additionally, we also examine the effect of hydrogen passivation of the surface defects considered.
The account of the vdW interactions allowed us to estimate the vdW part of the graphene-SiO$_2$ interaction, which is found to be ranging from 23.8 to 39.6 meV/C depending on the particular surface type. 
In the case of hydrogen passivated surfaces as well as for the silicon-terminated surface, the vdW interactions play a dominant role. By contrast, for the 
oxygen-terminated surface, the vdW part is responsible only for half of the total binding energy. The rest arises from the charge transfer and from resulting ionic interactions
between graphene and the underlying surface.
In addition, we show that in the absence of interfacial covalent bonding with the substrate, the electronic structure of graphene remains virtually unaltered, which may play an important role for practical 
applications. Finally, calculated interlayer distances and obtained charge inhomogeneity for graphene on SiO$_2$ are found to be in agreement with the data of atomic force microscopy (AFM) 
and STM experiments,\cite{Ishigami,Zhang} respectively. This agreement justifies the structural model and computational approaches used in the present study.

The paper is organized as follows. In Sec.~II, we briefly describe the methods of the investigation. In Sec.~III, we present the structural model of the SiO$_2$ surface and
describe other structural parameters used in our study. In Sec.~IV~A, we analyze the properties of the model SiO$_2$ surfaces. 
The results on interface geometry, adhesion, and electronic properties of graphene on SiO$_2$ are discussed in Secs.~IV~B, IV~C, and IV~D, respectively. Comparison with other theoretical studies are 
given in Sec.~IV~E. In the last section we briefly summarize our results.

\section{Calculation details}

The results presented in this study have been obtained based on the plane-wave 
pseudopotential method as implemented in the {\sc quantum-espresso} simulation
package.\cite{espresso,pseudo} 
To calculate adhesion energies and properly take into account dispersive
interactions, exchange and correlation effects
have been treated according to the vdW-DF method proposed by 
Dion \emph{et al}.\cite{Dion,Thonhauser} 
Up to now, this method has been extensively applied to a large variety of compounds
showing transferability across a broad spectrum of interactions, such as ionic, 
covalent, and vdW.\cite{Langreth} Although the vdW-DF method does not usually provide 
chemical accuracy, it has distinct advantages over the standard (semi)local approximations, 
such as explicit inclusion of the nonlocal electron correlations, leading to the proper
description of the dispersive interactions.

Within the vdW-DF, the exchange-correlation energy functional
consists of several contributions,
\begin{equation}
E_{xc}(n)=E_x^{\mathrm{revPBE}}(n)+E_c^{\mathrm{LDA}}(n)+E_c^{\mathrm{nl}}(n),
\label{exc}
\end{equation}
where the first term corresponds to the exchange part of the revised
PBE (revPBE) functional,\cite{Zhang-Yang} $E_c^{\mathrm{LDA}}$ is the
correlation energy within the LDA, and $E_c^{\mathrm{nl}}(n)$ is the nonlocal
correlation correction, which is calculated in the following way:
\begin{equation}
E_c^{\mathrm{nl}}=\frac{1}{2}\int d^3r d^3r' n(r)\phi(r,r')n(r'),
\label{nonlocal}
\end{equation}
where $n(r)$ is the electronic density and $\phi(r,r')$ is a
function incorporating many-body density response (for details see
Ref.~\onlinecite{Dion}).
It should be emphasized that
we evaluate the nonlocal correction [see Eq.~(\ref{nonlocal})] in a
perturbative way, i.e., using only GGA-based (semilocal)
electronic density distribution. This choice looks quite reasonable since 
the vdW interactions are rather weak and thus cannot significantly change the 
electronic distribution. Moreover, as has been shown on different vdW complexes,
the effects due to lack of self-consistency are negligible.\cite{Thonhauser}
To calculate this correction to the total energy, we used an efficient 
post-processing routine {\sc noloco} from the {\sc exciting} code.\cite{noloco,exciting}

Self-consistent treatment of the dispersion effects within the vdW-DF
approach is considerably more demanding in terms of the computational time. As a 
result, an efficient calculation of forces acting on atoms is also rather problematic
for large systems. For this reason, we do not use straightforward geometry optimization 
of the graphene/SiO$_2$ system within the vdW-DF functional. Instead, we optimize only 
the isolated SiO$_2$ surfaces within the gradient corrected functional in
the revised PBE parametrization.\cite{Zhang-Yang} This choice appears reasonable
since the vdW interactions can be neglected for the energies of silica structures.
Furthermore, we assume that the structure of graphene remains essentially unperturbed upon 
adhesion (see Sec.~IV~B for discussion). In order to find equilibrium separations 
between graphene and its substrates, we calculated adhesion energies as a function of the 
distance between them: 
\begin{equation}
E_{\mathrm{adh}}(d)=E_{\mathrm{Gr+Surf}}(d)-E_{\mathrm{Gr}}-E_{\mathrm{Surf}},
\end{equation}
where $E_{\mathrm{Gr+Surf}}(d)$ is the vdW-DF energy of graphene separated at the
distance $d$ from the surface, whereas $E_{\mathrm{Gr}}$ and
$E_{\mathrm{Surf}}$ are the energies of noninteracting graphene and the surface,
respectively. The minimum of the given function corresponds to the 
adhesion energy at the equilibrium separation, $\mathrm{min}\{E_{\mathrm{adh}}(d)$\}=$E_{\mathrm{adh}}(d_{\mathrm{eq}})$.

In our calculations, we employed an energy cutoff of 30 Ry for the
plane-wave basis and 300 Ry for the charge density.
Self-consistent calculations of the Kohn-Sham equations were carried out
employing the convergence criterion of 10$^{-8}$ Ry. For accurate
Brillouin-zone integration, the tetrahedron
scheme\cite{tetrahedron} and a (16 $\times$ 16 $\times$ 1)
Monkhorst-Pack {\bf k}-point mesh \cite{monkhorst} were used.
A much finer mesh (48 $\times$ 48 $\times$ 1) and a Gaussian broadening 
of 0.01 Ry were used for the density of states (DOS) calculations.
We checked that further increase of computational accuracy does not
significantly change the results.
In all cases under consideration, the height of the supercell was chosen 
to be 50 \AA. In order to avoid spurious interactions between images of the
supercell in the [001] direction, we also used a dipole correction.\cite{dc}
The convergence criterion for the relaxation of the SiO$_2$ surfaces was set 
to 0.001 Ry/\AA. The positions of the lowermost layer of atoms were fixed.

\section{Surface structures}

First-principles studies of amorphous compounds like SiO$_2$ are challenging
due to lack of translational symmetry in these systems. 
In many cases the surface structure of amorphous SiO$_2$ is modeled using various crystalline 
approximations.\cite{sio2_Wehling,Kang,Shemella,Li,Hossain,Jadaun,Cuong,vigne,ceresoli,ricci,jiang} 
Specifically, various faces and terminations of $\alpha$-quartz and $\beta$-cristobalite
are frequently used to reproduce the structure of SiO$_2$.
In this respect the structure of $\beta$-cristobalite seems to be more appropriate
since $\beta$-cristobalite and amorphous SiO$_2$ have very similar
local structures, as has been shown by neutron diffraction 
experiments.\cite{swainson,keen} However, in contrast to crystalline phases of 
SiO$_2$, the surface structure of the amorphous phase is nonuniform and has 
different reactive sites, which leads to unique chemistry of this
surface.\cite{walsh_2}

\begin{figure}[!tbp]
\includegraphics[width=0.50\textwidth, angle=0]{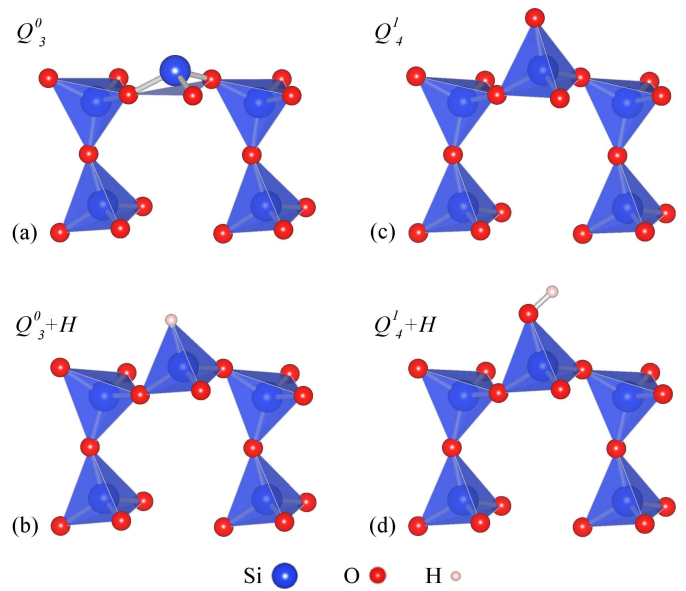}
\caption{(Color online) Surface structures of SiO$_2$ with the main types of defects 
(side view): (a) threefold coordinated silicon, $Q_3^0$, (b) $Q_3^0$ terminated by hydrogen atom, 
(c) single oxygen dangling bond, $Q_4^1$, and (d) $Q_4^1$ terminated by hydrogen atom.}
\label{sio2_str}
\end{figure}

According to atomistic molecular dynamics simulations,\cite{walsh_1} there 
are three types of stable point defects on amorphous SiO$_2$ surface: (i) O$_3$$\equiv$Si\textbullet, 
threefold coordinated silicon with no dangling oxygen bonds ($Q_3^0$),
(ii) O$_3$$\equiv$Si$-$O\textbullet, fourfold coordinated silicon with one dangling oxygen bond ($Q_4^1$), 
and (iii) threefold coordinated silicon with one dangling oxygen bond ($Q_3^1$). 
The notation $Q_m^n$ here implies $m$-coordinated silicon with $n$ oxygen
dangling bonds. For convenience, we will use this notation throughout the paper.
Moreover, as follows from the atomistic simulations, $Q_3^0$ and $Q_4^1$ are the most 
common defect sites ($\sim $80\% of the total surface area), which is in agreement with other theoretical 
investigations.\cite{Feuston} From the experimental point of view 
these two defect types also can be considered as fundamental in amorphous SiO$_2$.\cite{Messina}
An additional important type of chemical environment for
silica is $Q_4^0$, which corresponds to bulk silica without defects and
can be also modeled as a passivated surface defect (e.g., by hydrogen). 
In practice, partial passivation of the SiO$_2$ surface can be achieved by exposure to
molecular hydrogen (H$_2$) even at room temperatures.\cite{Messina}
In principle, the emergence of other surface defects is possible, but their 
existence is energetically unfavorable and, with high probability, leads to a local 
reconstruction of the surface. Interestingly, the reconstructed surface of crystalline 
$\alpha$-quartz phase is predicted to be fully saturated owing to the formation of six-membered 
rings at the surface.\cite{Rignanese,Goumans} This is, however, a consequence of the long-range 
order of crystalline surfaces and is not likely to occur for amorphous surfaces.

Unsaturated surfaces are usually reactive and easily adsorb impurities from the environment to
passivate themselves. Such a passivation can be, in principle, avoided under ultrahigh vacuum (UHV) conditions.
However, as has been previously demonstrated by EFM experiments, the amorphous SiO$_2$ surface shows relatively large fluctuations of electrical potential at ambient conditions.\cite{Garcia} 
This fluctuations suggest the existence of uncompensated surface charges even without UHV, which can be associated with unsaturated surface defects. 

\begin{figure}[t]
\includegraphics[width=0.50\textwidth, angle=0]{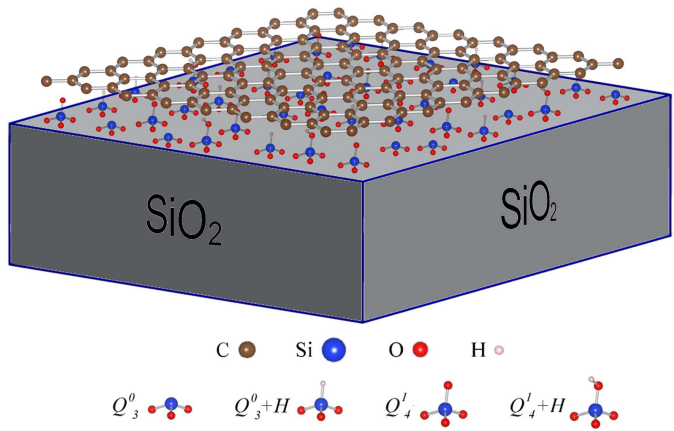}
\caption{(Color online) Schematic representation of graphene supported on a SiO$_2$ surface with different defects on the top (according to the model employing in this work).}
\label{sio2gr_str}
\end{figure}

We believe that the chemical reactivity of realistic amorphous silica can be
defined by a certain combination between the reactivities of local surface regions. This regions
can be modeled by idealized surfaces, each of which contains only one defect type among those discussed above. 
In turn, the properties of graphene at a large scale (realistic SiO$_2$) are expected to be composed of its properties
at the local scale (idealized SiO$_2$). Besides computational advantages within the first-principles framework, this 
approach allows us to investigate the role of different surface defects separately.
In our work, for modeling the SiO$_2$ surface, we 
consider two different $\beta$-cristobalite surfaces with morphologies 
corresponding to the $Q_3^0$ and $Q_4^1$ defect sites. We show relaxed structures 
of these surfaces in Figs.~\ref{sio2_str}(a) and \ref{sio2_str}(c). As we already
mentioned, these two defects are the most numerous types in amorphous
SiO$_2$ according to the previous studies.\cite{Feuston,walsh_1} 
We do not consider the $Q_3^1$ defect mainly because
it would have been necessary to consider a (100) surface of $\beta$-cristobalite  
instead of a (111) surface that is suitable for the modeling of $Q_3^0$ and $Q_4^1$-defective
surfaces. However, the unit cell of a (100) $\beta$-cristobalite surface is not
commensurate with the unit cell of graphene in lateral directions. In this case,
the lattice constant mismatch is more than 10\% and the corresponding adjustment of 
the SiO$_2$ unit cell would considerably affect the properties of the surface.
Since the $Q_3^1$ defect contains one oxygen dangling bond and one three-coordinated silicon, it is
reasonable to suppose that this defect type may be approximated by two separated defects, $Q_4^1$ and
$Q_3^0$.
In addition to unsaturated surfaces, we investigate the effect 
of hydrogen passivation on adhesive properties of SiO$_2$ surfaces [see Figs.~\ref{sio2_str}(b) 
and \ref{sio2_str}(d)].
In Fig.~\ref{sio2gr_str}, we show schematically the structure of graphene supported on SiO$_2$ with various defects on its surface.

In all cases, the supercell used in our study contains three layers of 
defectless (SiO$_4$)$^{4-}$ tetrahedra and one topmost defective layer. 
Oxygen atoms in the lowermost layer are saturated by hydrogen atoms.
All atoms except the lowermost layer of hydrogen were relaxed.
We used lateral $(2\times2)a$ graphene unit-cell parameters for the SiO$_2$-graphene supercell, 
which corresponds to a $\sim$3\% laterally compressed (111) surface unit cell of 
$\beta$-cristobalite.\cite{beta-crist} 
We believe this choice is justified since silicates are known to have a relative low bulk moduli, 
i.e., they can be compressed quite easily.\cite{min_handbook}
 
As for graphene, it has a two-dimensional honeycomb lattice of $sp^2$-bonded carbon atoms.
Although the real structure of graphene is corrugated, the characteristic
length of corresponding ripples is around 100 \AA,\cite{Meyer,Fasolino} which
is much larger than the typical length of supercells used in first-principles
calculations. For this reason, we do not consider this phenomenon in our work.
We used a lattice constant of graphene equal to $a$~=~2.459 \AA \
in accordance with the experimentally obtained value for graphite at
low temperatures.\cite{Baskin}

\section{Results and discussion}

\subsection{Bare SiO$_2$ surfaces}
\begin{figure}[!tbp]
\vspace{0.5cm}
\includegraphics[width=0.47\textwidth, angle=0]{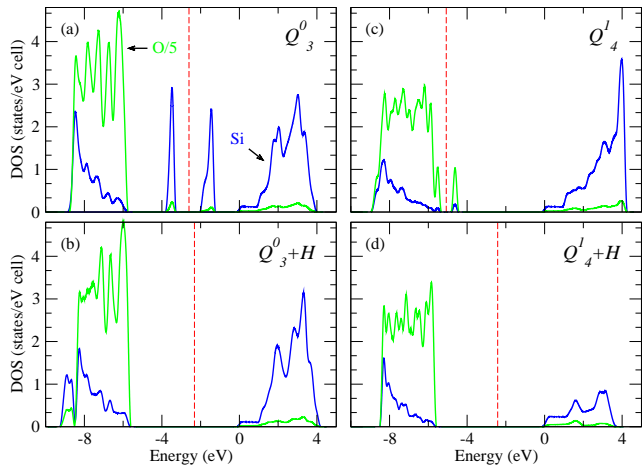}
\caption{(Color online) Densities of states for SiO$_2$ surfaces projected onto Si (blue line) and O (green line) atoms: (a) and (c) surfaces covered by the $Q_3^0$ and $Q_4^1$ defects, 
respectively; (b) and (d) corresponding hydrogen-passivated surfaces. Oxygen DOS is reduced by a factor of five for clarity. The vertical line corresponds to the Fermi level. Zero energy coincides with the lower energy of the conduction band.}
\label{sio2_dos}
\end{figure}

In Fig.~\ref{sio2_dos}, we show densities of electronic states for the SiO$_2$ surfaces under investigation. One can see that the H-passivated surfaces [see Figs.~\ref{sio2_dos}(b) and \ref{sio2_dos}(d)] 
are fully saturated and
there are no resonances within the band gap, indicating the absence of localized states at the surface. In these cases, the DOS exhibits a band gap of $\sim$5.7 eV, which is comparable with the band gap of 
bulk $\beta$-cristobalite calculated using the PBE 
functional (5.84 eV).\cite{Luppi} These values, however, underestimate
the experimental gap of amorphous SiO$_2$ films (9.3 eV),\cite{Weinberg} which is related to the well-known problem of the standard density functional theory (DFT) functionals (LDA/GGA), such as 
inability to accurately describe 
electronic spectra for compounds with strong electron localization. The band gap of silica can be considerably improved using hybrid functionals (e.g., PBE0)\cite{Alkauskas} or $GW$ 
approximation.\cite{Ramos}
In our work, however, we are not focused on a precise quantitative description of the electronic properties of SiO$_2$. We believe that the use of more appropriate
approaches will not qualitatively affect the results obtained in this study.

In contrast to H-passivated SiO$_2$, defective surfaces are unsaturated and, therefore the DOS exhibits resonances within the band gap arising from the localized surface states. 
For each surface, there are two peaks corresponding to occupied and unoccupied states of opposite spin.
In the case of
surface with undercoordinated silicon (the $Q_3^0$ defect), the midgap states originate from the Si($3s$,$3p$) and O($2p$) orbitals, whereas for the $Q_4^1$ defect the resonance is slightly more 
intense and localized mainly at the O($2p$) orbital. 
The presence of localized surface states allows one to expect chemisorption of graphene on a surface. Indeed, with respect to graphene supported on SiO$_2$, the defect sites may 
be effectively considered as monovalent impurities. In turn, such impurities may bind ionically or covalently with graphene.\cite{imp_Wehling}

    \begin{table}[!tbp]
    \centering
    \caption[Bset]{Work functions (WF) and electron affinities (EA) calculated for different SiO$_2$ surfaces (see Sec.~III for details) and graphene.}
    \label{sio2_tab}
\begin{ruledtabular}
 \begin{tabular}{ccccccc}
                      &   $Q_3^0$  & $Q_3^0+H$ & $Q_4^1$  & $Q_4^1+H$ & graphene \\
     \hline
  WF, eV            &    5.6    &    8.0   &   9.3   &    7.2   &   4.2   \\
  EA, eV            &    4.1    &    2.3   &   8.6   &    1.6   &   4.2   \\

    \end{tabular}
\end{ruledtabular}
    \end{table}

In order to estimate chemical activity of the surfaces in respect to graphene, we calculate the work functions (WF) and electron affinities (EA) as ($E_{\mathrm{vac}}$-$E_F$) and ($E_{\mathrm{vac}}$-$E_{\mathrm{cond}}$),
respectively, where $E_{\mathrm{vac}}$ is the vacuum level of the electrostatic potential, $E_F$ is the Fermi level, and $E_{\mathrm{cond}}$ is the lowest energy of the conduction band.
We stress, however, that the accuracy of the WF and EA is limited within the employed 
computational method, similar to the band-gap problem (see above). Nevertheless, these 
quantities are useful to establish general trends in the chemisorption of graphene on SiO$_2$.
In Table \ref{sio2_tab}, we summarize calculated WF and
EA for SiO$_2$ surfaces and graphene. For graphene, WF is equal to EA due to the absence of a band gap. Comparing WF and EA for graphene and H-passivated surfaces, one can see that the 
surface WF is larger than the EA of graphene (and vice versa). The same situation occurs for graphene and the surface with the 
$Q_3^0$ defects. This implies that the 
charge transfer between these surfaces is not energetically favorable. On the other hand, the WF of graphene is lower than the EA of the surface with oxygen dangling bond ($Q_4^1$), which suggests 
electronic transfer from graphene toward this surface ($p$-type doping of graphene).

\subsection{Geometry of the graphene-SiO$_2$ interface}
As we already mentioned in Sec.~II, we calculate equilibrium graphene-SiO$_2$ distances assuming that the structure of graphene is unperturbed upon contact with the substrate. However, this
does not reflect the structure of graphene on SiO$_2$ on a large scale. 
As follows from experimental studies, the structure of graphene supported on amorphous SiO$_2$ is highly corrugated owing to high-fidelity conformation.\cite{Geringer,Ishigami,Cullen} 
The assumed invariability of graphene is resulting from the employed theoretical approach. Indeed, the supercell used in our study is too small in lateral directions to reveal 
corrugations of graphene caused by the irregularity of the substrate. Moreover, imposed 
boundary conditions along with the strong $sp^2$ bonding of carbon atoms artificially restrict the flexibility of graphene. Previous results show that for such small supercells the deformation of
graphene is really negligible.\cite{mica} Thus we note that our results reflect only the local 
behavior of graphene on SiO$_2$. Therefore the interlayer distance is the main structural parameter in our model. In Table \ref{sio2gr_tab}, we summarize 
these parameters for different SiO$_2$ surfaces.

Equilibrium distances between graphene and SiO$_2$ are considerably different for the surfaces covered by the $Q_3^0$ and $Q_4^1$ 
defects. This can be easily accounted for by 
noticing that the difference in the distance correlates with the difference in the adhesion energy. Namely, the smaller distance corresponds to the larger $E_{\mathrm{adh}}$ and vice versa. 
The situation for hydrogen-passivated defects is different. In this case, although the binding energy is smaller relative to the binding for the $Q_3^0$ defect, the interlayer 
distance is also smaller. 
This is caused by the fact that the orbitals associated with hydrogen have a small spread, i.e., 
they are not much spatially extended compared to the oxygen or silicon orbitals. As a consequence, 
the Pauli repulsion between graphene and the hydrogen layer decreases more rapidly with distance, 
which allows graphene to come closer to hydrogen.

    \begin{table}[!bt]
    \centering
    \caption[Bset]{Calculated interlayer distances, adhesion energies, Fermi level shifts, and WFs (EAs) for graphene supported on different SiO$_2$ surfaces. 
Interlayer distances are calculated as a difference between the $z$ coordinate of carbon atoms in graphene and the topmost atomic layer of the substrate. The Fermi level shift implies the difference
between the Dirac point and the Fermi level of supported graphene. The values are given for the most stable graphene geometry.}
    \label{sio2gr_tab}
\begin{ruledtabular}
 \begin{tabular}{cccccc}
                         &   $Q_3^0$  & $Q_3^0+H$  & $Q_4^1$  & $Q_4^1+H$ \\
     \hline
  $d_{\mathrm{eq}}$, \AA               &    3.6     &    2.7     &    2.7    &    2.6   \\
  $E_{\mathrm{adh}}$, meV/C       &   $-$31.6   &   $-$26.5   &  $-$77.7  &  $-$24.8  \\
  vdW part of $E_{\mathrm{adh}}$\footnotemark[1]  &     98\%   &     96\%   &    51\%   &    96\%  \\
  $\Delta E_F$, eV       &    0.0    &     0.0    &    $-$1.0   &    0.0   \\
  WF (EA), eV       &    4.2    &   4.2     &    5.2   &    4.2   \\

    \end{tabular}
\end{ruledtabular}
\footnotetext[1]{Contribution of vdW interaction calculated as a difference between adhesion energies
in the presence of nonlocal correlations and without it.}
    \end{table}

The interface distances obtained between graphene and SiO$_2$ can be compared with experimental height histograms acquired across the graphene-SiO$_2$ boundary
using the AFM technique.\cite{Ishigami} According to the experimental data, the average thickness of a single graphene layer deposited on SiO$_2$ under UHV
conditions can be estimated as 4.2 \AA. This quantity, however, does not reflect the actual distance of the interface since it directly depends on the spatial extension of the 
orbitals of both surfaces. In fact, the $\pi$ orbitals of graphene are known to have a rather large spread (see, e.g., Ref.~\onlinecite{Freimuth}) and, therefore, the thickness of 
graphene supported on a surface with more localized orbitals should be larger than the distance in the theoretical sense, i.e., as a difference between atomic coordinates. 
On the other hand, if the thickness of the topmost atomic layer is \emph{much} smaller than the thickness of supported graphene, the interlayer distance corresponds to  
half of the graphene thickness. It is difficult to establish a more precise relationship between these quantities because of a number of ambiguities in the
determination of the orbital spread and its quantitative influence on experimental results. According to this consideration, the average experimental distance should
satisfy the following uncertainty relation: $h/2 < d_{\mathrm{exp}} < h$, where $h$ is the average experimental thickness. Thus, for the mentioned experiment,\cite{Ishigami} the 
distance $d_{\mathrm{exp}}$ is bounded within 2.1\---4.2 \AA. 
One can see that obtained theoretical distances calculated according to the given structural model of the SiO$_2$ surface (see Table \ref{sio2gr_tab}) fall entirely within this range.
Taking the uncertainty into account, we come to a good agreement between results obtained and experimental estimations. 
It should be noted that the obtained distances are rather sensitive to the presence of the vdW interactions, especially if they dominate. In fact, the distances calculated without the nonlocal term in
the exchange-correlation functional [see Eq.~(\ref{nonlocal})] are significantly higher and therefore do not lead to an agreement with the experiment.

\subsection{Graphene adhesion on SiO$_2$}
As follows from Table \ref{sio2gr_tab}, the main contribution (96\%) to adhesion of graphene on H-passivated SiO$_2$ surfaces 
comes from the vdW interaction. Adhesion energies for $Q_3^0+H$ and $Q_4^1+H$ cases are very close to each other, which indicates that 
the vdW part of the adhesion energy is only slightly dependent on the particular surface type. The situation for unsaturated surfaces is 
partially different.
In the case of a surface with undercoordinated silicon ($Q_3^0$), the vdW contribution still dominates (98\%) leading to the adhesion energy
that is slightly larger than for the H-passivated case, whereas the presence of dangling oxygen bonds on the surface ($Q_4^1$) 
results in a much stronger adhesion of only 51\% of the vdW interaction. 
The latter implies that graphene definitely chemisorbs on the surface with $Q_4^1$ defect, which is the most reactive among 
the others.
The values in Table \ref{sio2gr_tab} are given for the most stable configuration of graphene relative to the considered surfaces. Namely, for all considered surface defects, this configuration
corresponds to the situation where point defects are situated below a hollow (hexagon center) graphene site. If the defect is located below a bridge (center of the C-C bond) or top site (C atom),
the maximum difference in adsorption energy is reached for the oxygen terminated surface ($\sim$5 meV/C). For three other surfaces considered, this difference is not exceeding 1 meV/C.
It is worth mentioning that irrespective of the surface type, values obtained for the adhesion energies are sufficiently exceeding the elastic energy stored in 
graphene ($\sim$0.8\---2.6 meV/C).\cite{Ishigami,Cullen} 
This means that the interaction energy between graphene and SiO$_2$ is large enough to overcome the energy needed for graphene to follow the SiO$_2$ morphology. Therefore, our results 
confirm previous estimations regarding the possibility of graphene-SiO$_2$ conformation.

Very recently, the first direct measurements of the adhesion energy of 1\---5 layer graphene to SiO$_2$ were reported.\cite{Koenig} For monolayer graphene the value 
of $0.45\pm0.02$ J/m$^2$ ($\sim74\pm3$ meV/C) 
was obtained. As can be seen, the experimental adhesion energy is very close to our theoretical estimation (77.7 meV/C) for the oxygen-terminated surface ($Q_4^1$ defect), but significantly higher than 
for the other surface types. This might imply that the $Q_4^1$ defect is the most common defect at the SiO$_2$ surface in the experiment. 
However, in view of the fact that the $Q_4^1$ defect can be considered as a charge impurity (as we show below), this assumption seems to be inconsistent with the reported expectations that charge 
impurities should not significantly affect the adhesion because of their relatively low density in SiO$_2$. We believe that in order to resolve this contradiction, a more reliable large-scale model for 
the SiO$_2$ structure is required, where concentration and distribution of the (sub)surface defects are explicitly taken into account. We leave this problem open for 
further studies.

\subsection{Electronic structure and charge transfer in supported graphene} 
\begin{figure}[!tbp]
\includegraphics[width=0.45\textwidth, angle=0]{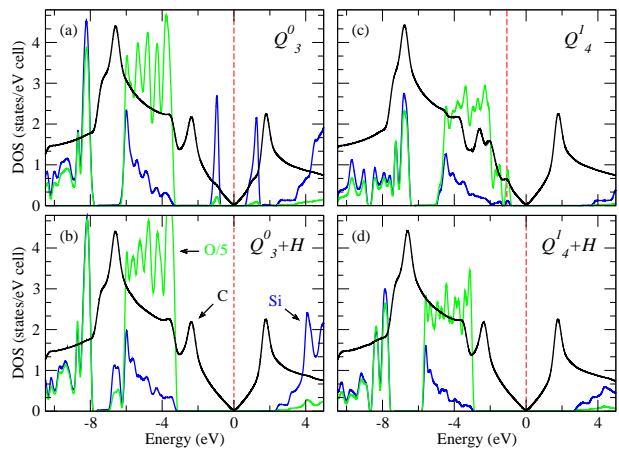}
\caption{(Color online) Densities of states of SiO$_2$ surfaces in contact with graphene sheet projected onto C (black line), Si (blue line), and O (green line) atoms. 
The SiO$_2$ substrates are given in the same order as in Fig.~\ref{sio2_dos}. Oxygen DOS is reduced by a factor of five. The vertical line corresponds to the Fermi level. Zero energy coincides with the Dirac point of graphene.}
\label{sio2gr_dos}
\end{figure}

In Fig.~\ref{sio2gr_dos}, we show DOS for graphene on SiO$_2$. As there is no chemisorption between graphene and H-passivated SiO$_2$ surfaces, the electronic structure of these systems is not changing
upon adhesion [see Figs.~\ref{sio2gr_dos}(b) and \ref{sio2gr_dos}(d)]. In this situation, the Dirac point of graphene coincides with the Fermi level and lies in the center of the substrate band gap. 
The same is true for the 
$Q_3^0$-defective surface with the exception that the DOS of graphene is slightly distorted in the vicinity of impurity resonances. This may indicate a slight orbital hybridization between graphene and the
surface. However, these distortions are small enough to produce any noticeable changes in the adhesion energy.
As we have shown before, in the case of graphene adherent to the $Q_4^1$ surface, there should be additional sources of the interaction apart from the vdW one. Namely, as follows from Table \ref{sio2_tab}, 
$p$-type doping of graphene is expected since the EA of the $Q_4^1$ surface is larger than the WF of isolated graphene. This is also clear from the DOS [see Fig.~\ref{sio2gr_dos}(c)] that exhibits the 
Fermi 
level shift toward the low-energy range relative to the Dirac point. Such a shift means that graphene works as a donor on the $Q_4^1$-defective SiO$_2$ surface.
As can be also seen in Fig.~\ref{sio2gr_dos}(c), the Fermi level lies at the impurity resonance suggesting that the electrons from graphene are transferred to the $p$-orbital of oxygen, which becomes 
partially occupied. L\"{o}wdin charge analysis\cite{Lowdin} shows the charge doping of 0.29 $e^{-}$/cell.

The Fermi level shift implies the existence of induced charges in local regions of supported graphene, which can be considered as Coulomb impurities. This finding is in agreement with experimental
STM studies, where electron-density inhomogeneities have been clearly observed in graphene flakes placed on a SiO$_2$ substrate.\cite{Zhang} 
Raman spectroscopy studies also provide some evidences for the $p$-type doping of graphene on SiO$_2$.\cite{Ryu}
Our results allow one to identify more specifically the origin of the charge inhomogeneities and associate them with the $Q_4^1$ surface defect. The presence of these impurities plays a significant role for
charge-carrier scattering and may affect the electron mobility of a sample. This is especially important for graphene supported on a SiO$_2$ substrate, where mobility is considerably limited
in comparison to suspended graphene.\cite{Chen} Although there are a number of sources restricting the mobility of graphene, charged impurities seem to be one of the dominating factors, at least at low 
temperatures.\cite{Chen2} It is interesting to note that the overall mobility of graphene depends on the distribution of the charged impurities on its surface.\cite{Cluster1,Cluster2} 
In particular, the contribution of impurities to the resistivity is maximal for their homogeneous distribution and can be strongly suppressed by clusterization. In the case under investigation, these
effects depend on the particular SiO$_2$ morphology.

Apart from the Fermi level shift indicating the charge transfer from graphene to the $Q_4^1$ surface, there is a distortion of the graphene DOS within the energy range of $-$3 to 1 eV relative
to the Dirac point. It is worth mentioning that no distortions occur in the vicinity of the Dirac point of graphene.
The alteration of the DOS reflects some hybridization between the $\pi$ orbitals of graphene and $p$ orbitals of oxygen from the 
defective layer of SiO$_2$. Since the oxygen atom has only one unsaturated electron in 
the $Q_4^1$ defect, this defect site can be considered as a monovalent impurity in respect to graphene. According to Ref.~\onlinecite{imp_Wehling}, monovalent adsorbates on graphene 
can be divided into two separate groups regarding the bonding mechanism: ionically and covalently bonded impurities. The main qualitative difference between these two groups is that the second type
of impurities induces a band gap and corresponding midgap states at the Fermi level, whereas the first type gives rise only to impurity states lying below (or above) the Dirac point. Importantly, the 
typical band structure of graphene is preserved for ionic group of impurities. In accordance with this classification,
the $Q_4^1$ defect site corresponds to an ionic impurity since it does not disturb significantly the band structure of graphene. We note that no qualitative changes in terms of the total DOS have been found 
among the different surface geometries of graphene.

Besides the covalent bonding, there is another mechanism of the band-gap opening in graphene, resulting from the symmetry breaking. If a binding partner interacts with one sublattice
of graphene more strongly than with another sublattice, the graphene sublattice symmetry is broken, resulting in the band-gap opening. One of the prominent examples exhibiting this effect is graphene
supported on the surface of hexagonal boron nitride.\cite{Giovannetti} For graphene supported on the SiO$_2$ surfaces, we also found this effect, however, in all cases considered the interaction between 
graphene and its substrate is relatively weak and the band gap is not exceeding 15 meV. Since this value is less than the thermal energy at room temperature ($\sim$25 meV), 
we believe that this gap can be considered as negligible.

In Table \ref{sio2gr_tab}, we also give WFs of supported graphene. Since the substrate does not induce a band gap in graphene, its WFs coincide with EAs, as for isolated graphene. One can see that
for all considered substrates, with the exception of the oxygen-terminated SiO$_2$ surface, the WF of supported graphene is equal to the WF of isolated graphene. This is a consequence of the absence of
the chemical bonding between graphene and these surfaces. The WF (EA) of graphene on the surface with $Q_4^1$ defects is larger by the value of the Fermi level shift caused by the charge transfer.
On a large scale, this means that the graphene sheet deposited on amorphous SiO$_2$ should possess unequal reactivity for different surface regions. This property might find application in practice
as a novel graphene-based method for microscopic surface studies, along with existing ones.\cite{Xu}

\subsection{Comparison with other theoretical studies} 
Our results, in particular concerning the oxygen-terminated surface ($Q_4^1$ defect), might appear qualitatively inconsistent with previous DFT investigations of graphene on SiO$_2$ 
surfaces.\cite{Kang,Shemella,Jadaun} In these works,
an evident covalent bonding has been observed between graphene and SiO$_2$ along with significant structural reconstruction both of graphene and the surface. Moreover, according to these investigations 
the electronic 
structure of graphene undergoes significant modification upon contact with SiO$_2$. Considerable band-gap opening is also reported. It should be noted that the results 
mentioned have been
obtained using the structure of $\alpha$-quartz for modeling the SiO$_2$ surface. This means that a cut of this structure in the [001] direction corresponds to a surface with two oxygen dangling bonds 
per unit cell or, in other words, to a $Q_4^2$-defective surface. In contrast, our results for the oxygen-terminated surface have been obtained using the surface with only one oxygen dangling 
bond ($Q_4^1$ defect) per unit
cell. Therefore, the mentioned inconsistency is not surprising since the $Q_4^2$ defect is apparently more reactive than the $Q_4^1$. This is also consistent with Ref.~\onlinecite{Hossain} where no
strong binding with graphene has been found in the case of $Q_4^1$-like surface termination. As we already stated in Sec.~III, the existence of free-standing
surface configuration like $Q_4^2$ is not energetically favorable and most probably would lead to a local reconstruction. 
Interestingly, if graphene is supported on a fully reconstructed SiO$_2$ surface,\cite{Cuong} neither covalent nor ionic interfacial interactions occur, which means that graphene 
\emph{physisorbs}
on the surface. The qualitative difference in the properties of graphene on various SiO$_2$ surfaces accentuates the importance of the realistic surface morphology in the determination
of graphene properties.
The results of electronic structure calculations for the hydrogenated surfaces considered in 
our work are in qualitative agreement with the mentioned investigations. Additionally, we note that the previous results were mainly obtained using the LDA functionals and, therefore, some quantitative 
discrepancies (e.g., in binding energies or interface distances) are legitimate.

\quad

\section{Conclusion}

We have investigated the nature of interfacial interactions between graphene and amorphous SiO$_2$ from first principles. We modeled SiO$_2$ surface assuming
the existence of certain types of surface defects that are relevant for amorphous silica surface.
Particularly, idealized surfaces with the oxygen dangling bond and three-coordinated silicon defects are considered, as well as the case of their hydrogen passivation.
We found that the vdW interaction plays a dominant role in the adhesion of graphene on silicon- and hydrogen-terminated surfaces. This confirms the importance of the dispersive
interactions for graphene supported on surfaces and also justifies the use of the nonlocal vdW correction. For the surface with dangling oxygen bonds, we found, apart from the vdW, an additional 
ionic contribution to the graphene-SiO$_2$ binding resulting from the electron transfer from graphene to the more electronegative SiO$_2$ surface. This contribution is responsible for half of
the total binding energy. 

We found that the DOS of supported graphene remains almost unperturbed retaining its typical Dirac shape. As a consequence, there is no covalent bonding between graphene and the surfaces considered.
The only change in the electronic structure of graphene concerns the Fermi level shift for the oxygen-terminated substrate. This shift, caused by the $Q_4^1$ defect, implies the existence of charged 
impurities and may be partially responsible for the known transport anomalies of graphene on SiO$_2$.

Our results show that different surface terminations lead to the variations of the distances between graphene and SiO$_2$. Unfortunately, these variations cannot be directly related to experimental 
observations due to
the complexity of the realistic SiO$_2$ morphology. Nevertheless, obtained interface distances are comparable with the average thickness of supported graphene observed in AFM experiments. This thickness
allows us to estimate the average graphene-SiO$_2$ distance. Taking into account ambiguities in the accurate determination of this distance, our theoretical results agree well with 
experimental data. This agreement shows the rationality of the structural model for amorphous SiO$_2$ used in our study.

\quad

\section{Acknowledgments}
We thank Tim Wehling and Sunmin Ryu for helpful discussions.
The authors acknowledge support from the Cluster of Excellence
``Nanospintronics'' (Hamburg, Germany), from Stichting voor
Fundamenteel Onderzoek der Materie (FOM, the Netherlands),
and from the Russian scientific programs Nos. 02.740.11.0217 and 2.1.1/779.
Figures \ref{sio2_str} and \ref{sio2gr_str} were generated using the {\sc vesta} program.\cite{vesta}

\end{document}